\documentclass[apj]{emulateapj}
\usepackage{apjfonts}

\shorttitle{The galactic spin of AGN galaxies}
\shortauthors{Cervantes-Sodi et al.}

\begin{document}

\title{The galactic spin of AGN galaxies}

\author{Bernardo Cervantes-Sodi \altaffilmark{1}, X. Hernandez \altaffilmark{2}, 
Changbom Park \altaffilmark{3} and Yun-Young Choi \altaffilmark{4}}
\altaffiltext{1}{Korea Astronomy and Space Science Institute, 61-1 Hwaam-dong, Yuseong-gu, Daejeon 305-348, Republic of Korea, bcsodi@kasi.re.kr}
\altaffiltext{2}{Instituto de Astronom\'\i a,
Universidad Nacional Aut\'onoma de M\'exico
A. P. 70--264,  M\'exico 04510 D.F., M\'exico,  xavier@astroscu.unam.mx}
\altaffiltext{3}{Korea Institute for Advanced Study, Dongdaemun-gu, Seoul 130-722, Republic of Korea, cbp@kias.re.kr}
\altaffiltext{4}{Department of Astronomy \& Space Science, Kyung Hee University, Gyeonggi 446-701, Republic of Korea}

\begin{abstract}

Using an extensive sample of galaxies selected from the Sloan Digital
Sky Survey Data Release 5, we compare the angular momentum distribution
of active galactic nucleus (AGN) with non-AGN hosting late-type galaxies. To this end we
characterize galactic spin through the dimensionless angular momentum parameter
$\lambda$, which we estimate approximately through simple dynamical considerations.
Using a volume limited sample, we find a considerable difference when comparing the empirical
distributions of $\lambda$ for AGNs and non-AGN galaxies, the AGNs showing
typically low $\lambda$ values and associated dispersions, while non-AGNs
present higher $\lambda$ values and a broader distribution. A more striking
difference is found when looking at $\lambda$ distributions in thin
$M_{ r}$ cuts, while the spin of non-AGN galaxies presents an anti-correlation
with $M_{ r}$, with bright (massive) galaxies having low spins,
AGN host galaxies present uniform values of $\lambda$
at all magnitudes, a behavior probably imposed by the fact that most late-type
AGN galaxies present a narrow range in color, with a typical constant $\lambda$ value.
We also find that the fraction of AGN
hosting galaxies in our sample strongly depends on galactic spin, increasing
dramatically for decreasing $\lambda$. For AGN host galaxies we compute the mass
of their supermassive black holes and find that this value tends to be higher
for low spin galaxies, even at fixed luminosity, a result that could account,
to a certain extent, for the spread on the luminosity black-hole mass relation.

\end{abstract}


\section{Introduction}

Studies of central gas and stellar kinematics of nearby galaxies have
revealed the presence of supermassive black holes (SMBHs) in almost all
massive galaxies. According to the most accepted paradigm (Rees 1984),
galactic nuclear activity is directly related with
the fueling process of these SMBHs, which strongly correlates to the overall
structure and evolution of the galaxies. For instance, active galactic nucleus (AGN) activity seems
to be triggered only in galaxies with massive bulges and intermediate
colors, as pointed out by Choi et al. (2009), henceforth CWP09.
These authors
show how the necessary conditions for AGN activity depend on galaxy
morphology; for early-type galaxies low velocity dispersion and
blue colors are preferred, and indicative of available gas supply for
the central black hole, while intermediate velocity dispersion,
intermediate-color, and more concentrated late-type galaxies are
more likely to host AGNs, reinforcing the idea that AGN host galaxies
are intermediate objects between the red sequence and the blue cloud
(Kauffmann et al. 2003).

If most galaxies present SMBHs in their centers,
a fundamental question is why not all of them host AGNs
when gas supply is available.
This is a question with no simple answer given the
complexity of the processes involved in the formation and evolution
of galaxies, but important clues have been proposed to elucidate the
mechanisms that trigger AGN activity (Martini 2004). Major mergers of gas-rich
progenitors and galaxy interactions are two commonly invoked mechanisms
that can provide gas to the center of the galaxies and fuel dormant
SMBHs (Hopkins et al. 2006; Ellison et al. 2008). Tidal torques
exerted by non-axisymmetric features such as large-scale and nuclear
bars (Jogee 2006), recycled gas from dying stars in the inner
kiloparsecs of the galaxy (Ciotti \& Ostriker 2007), and enhanced viscosity
of the turbulence associated with supernova explosions (Chen et al. 2009) are just
some other mechanisms that can effectively trigger nuclear activity.

A common requirement for any of the mentioned mechanisms is the removal of
specific angular momentum in order for the material to collapse to the
center, and operate as fuel. Therefore, a relation between the overall
galactic $\lambda$ spin parameter and the onset of AGN activity is expected,
with a higher frequency of AGNs in low spinning galaxies. Statistically,
AGN hosting galaxies should form a low $\lambda$ population, in comparison
to normal galaxies. Low angular momentum is expected not only in the
growing phase of SMBHs, but has also been invoked to explain the formation
of massive black hole seeds themselves. In this scenario
(Eisenstein \& Loeb 1995; Koushiappas et al. 2004), the direct
collapse of low angular momentum material results in a typical inverse
proportionality between the SMBH mass and the total angular momentum
(Haehnelt et al. 1998; Colgate et al. 2003).

In this work, we study the spin distribution
of AGN host galaxies in contrast with non-AGNs, parameterizing the
angular momentum through the dimensionless spin parameter

\begin{equation}
\label{Lamdef}
\lambda = \frac{L \mid E \mid^{1/2}}{G M^{5/2}},
\end{equation}

where $E$, $M$ and $L$ are the total energy, mass and angular momentum of the configuration, 
respectively (Peebles 1969). Our aim is to determine empirically if the AGN population
effectively presents a low spin distribution when compared with non-AGNs
and corroborate the results arising
from theoretical studies concerning the relation between $\lambda$ and the SMBH mass.
To this end, we use a sample of galaxies selected from the Sloan
Digital Sky Survey (SDSS) and apply a simple model to obtain
an empirical estimation of $\lambda$. The large volume limited sample
allows a thorough coverage of parameter space, which permits an empirical
determination of the leading galactic physical parameters driving the AGN
phenomena and the observed scalings with SMBH mass. The physical model is
presented in \S~2, and the sample selection criteria in \S~3. In
\S~4 we present our general results and the final conclusions
are summarized in \S~5.

\section{Estimation of the spin from observable parameters}

In Hernandez \& Cervantes-Sodi (2006), we derived a simple estimate of $\lambda$
for disk galaxies in terms of observational parameters
and showed some clear correlations between this parameter
and type defining structural parameters.
Here we briefly recall the main ingredients of the simple
model. The model considers only two components, a disk for
the baryonic component with an exponential surface mass density $\Sigma(r)$,

\begin{equation}
\label{Expprof}
\Sigma(r)=\Sigma_{0} e^{-r/R_{\rm d}},
\end{equation}

where $r$ is a radial coordinate, and $\Sigma_{\rm 0}$
and $R_{\rm d}$ are two constants which are allowed to
vary from galaxy to galaxy, and a dark matter halo having
a singular isothermal density profile $\rho(r)$,
responsible for establishing a rigorously flat rotation
curve $V_{\rm d}$ throughout the entire galaxy:

\begin{equation}
\label{RhoHalo}
\rho(r)={{1}\over{4 \pi G}}  \left( {{V_{\rm d}}\over{r}} \right)^{2}. 
\end{equation}

In this model we are further assuming that
(1) the specific angular momentum of the disk and halo are equal
(e.g. Mo et al. 1998; Zavala et al. 2008); (2) the total energy is dominated
by that of the halo which is a virialized gravitational structure; and
(3) the disk mass is a constant fraction of the halo mass,
$F=M_{\rm d}/M_{\rm H}$. These assumptions allow us to express $\lambda$ as

\begin{equation}
\label{Lamhalo}
\lambda=\frac{2^{1/2} V_{\rm d}^{2} R_{\rm d}}{G M_{H}}.
\end{equation}

Finally, we introduce a baryonic Tully--Fisher (TF) relation:
$M_{\rm d}=A_{\rm TF} V_{\rm d}^{3.5}$ (e.g., Gurovich et al. 2004).
Taking the Milky
Way as a representative example, we evaluate $F$ and $A_{\rm TF}$ to obtain

\begin{equation}
\label{LamObs}
\lambda=21.8 \frac{R_{\rm d}/\rm kpc}{(V_{\rm d}/\rm km \: s^{-1})^{3/2}}.
\end{equation}

We proved the accuracy in our estimation of $\lambda$ in Cervantes-Sodi et al. (2008),
comparing the estimation using
Equation (\ref{LamObs}) to values arising from numerical simulations of six distinct groups,
where the actual value of $\lambda$ is known, as it is one of the parameters
of the simulated galaxies, and where this can also be estimated through Equation (\ref{LamObs}),
as baryonic disk scale lengths and disk rotation velocities are part of the output.
The result was a one-to-one correlation, with very small dispersion and no bias,
leading to typical errors $<$ 30\%, which we include throughout.
The use of this estimate for large SDSS samples
led us in Hernandez et al. (2007) and Cervantes-Sodi et al. (2008) to a first empirical
derivation of the distribution of $\lambda$ parameters in the local universe
and a measure of the scalings of this parameter with local environment and halo mass.
These results were latter confirmed by Berta et al. (2008) using an estimate of
$\lambda$ through Equation (\ref{LamObs}), also used by Gogarten et al. (2010)
as an unbiased first-order $\lambda$ measurement.

\section{The SDSS sample}

The sample of galaxies used in this work is an enhanced version
of the study by CWP09. 
The galaxies come from the large-scale structure
sample, DR4plus, using the New York University
Value-Added Galaxy Catalogue (VAGC; Blanton et al. 2005), which is a galaxy subset
of the SDSS Data Release 5 (Adelman-McCarthy et al. 2006), and the
Korea Institute for Advanced Study-VAGC (Choi et al. 2010). A detailed
description of the sample can be found in Choi et al. (2007) and CWP09.
The rest-frame absolute magnitudes of individual galaxies were computed in
the $r$ band, using Galaxy reddening correction (Schlegel et al. 1998) and
$K$-corrections as described by Blanton et al. (2003), with a mean evolution
correction given by Tegmark et al. (2004).
We define a volume-limited sample with a lower magnitude cut-off
$M_{r} = -19.0$, with redshifts $0.020 < z < 0.086$, and a total number of
87,590 galaxies. Disk scale lengths are estimated from one-component light fits,
which implicitly account for the presence of low angular momentum material
in bulges.
 
\begin{figure}
\centering
\begin{tabular}{c}
\includegraphics[width=0.475\textwidth]{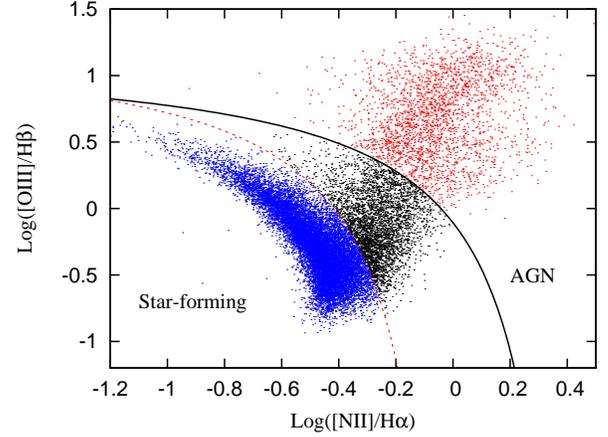}  
\end{tabular}
\caption[ ]{Galaxy distribution in the BPT diagram, pure AGN galaxies lying above the solid
line (Kewley et al. 2006), star-forming galaxies below the dashed line (Kauffmann et al. 2003),
and composite objects in between.}
\label{Fig1}
\end{figure}

In order to compare the spin of AGN and non-AGN galaxies, we use only Type II
AGNs to avoid complications due to the presence of a bright nucleus. The AGN
galaxies were segregated from star forming galaxies using the flux ratio
BPT diagram of Balmer and ionization lines (Baldwin et al. 1981).
The AGNs were selected among galaxies fulfilling the definition by
Kewley et al. (2006):

\begin{figure*}
\label{distributions}
\centering
\begin{tabular}{cc}
\includegraphics[width=0.475\textwidth]{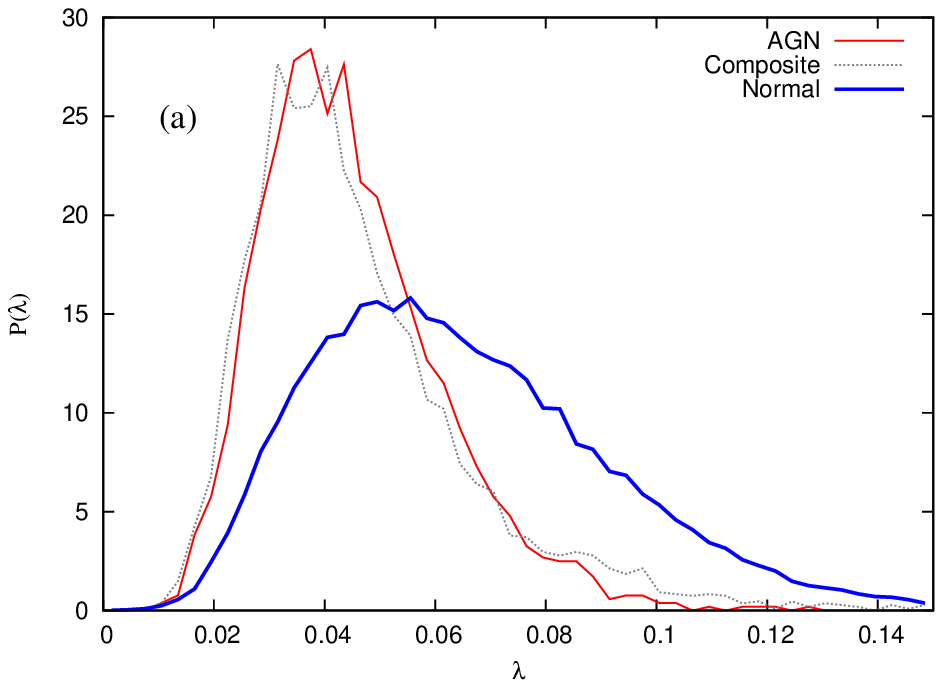} & \includegraphics[width=0.475\textwidth]{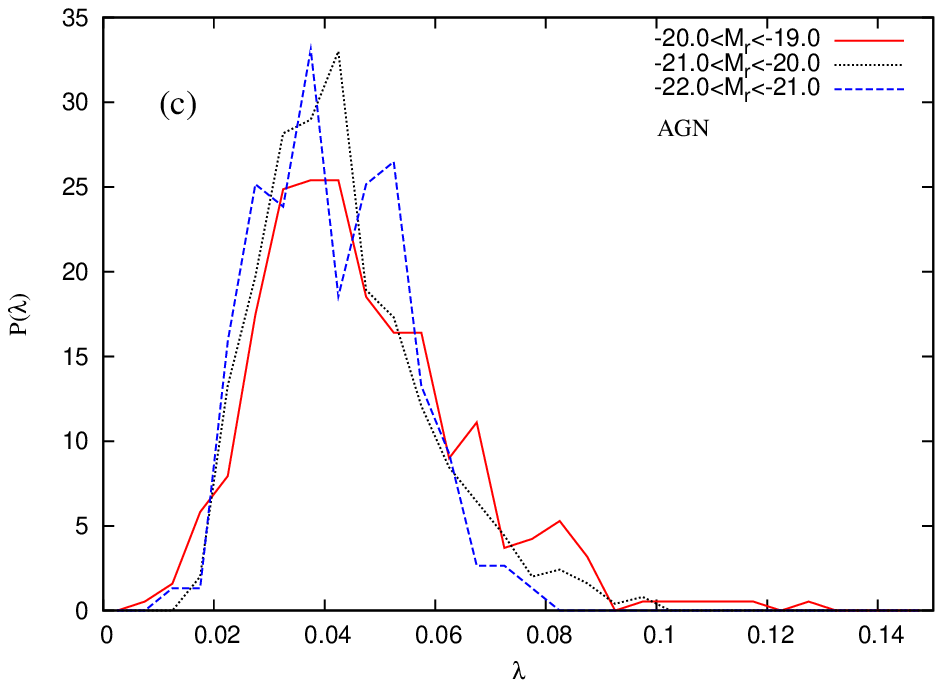} \\
\includegraphics[width=0.475\textwidth]{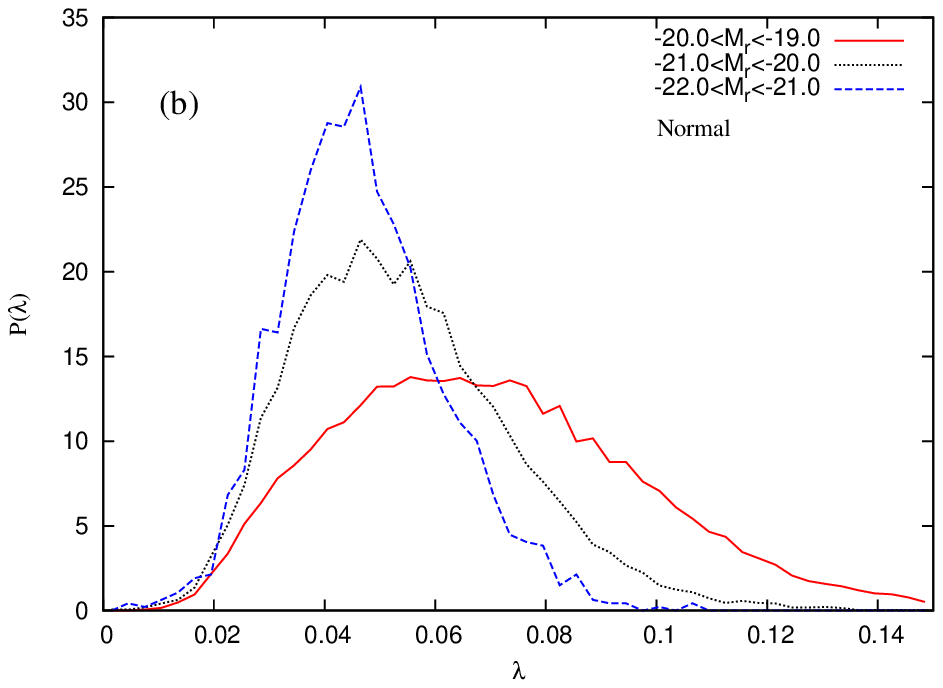} & \includegraphics[width=0.475\textwidth]{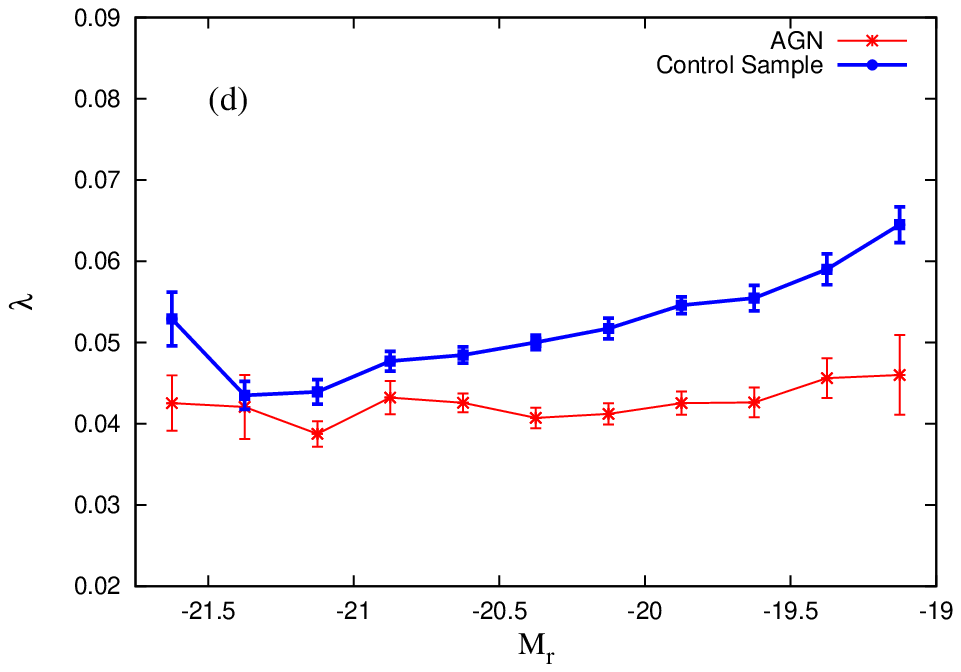}
\end{tabular}
\caption[ ]{(a) $P(\lambda)$ distribution for late-type
galaxies in the sample segregated in normal,
composite and AGN galaxies. (b) $P(\lambda)$ distribution
for normal galaxies split according to absolute magnitude.
(c) $P(\lambda)$ distribution
for AGN galaxies split according to absolute magnitude.
(d) Median $\lambda$ value of AGN galaxies and the control sample
as a function of $M_{\rm r}$.}
\end{figure*}

\begin{eqnarray}
0.61/({\rm log}([\rm N_{II}]/H\alpha)-0.47)+1.19&<&{\rm log}([\rm O_{III}]/H\beta),\nonumber\\
0.72/({\rm log}([\rm S_{II}]/H\alpha)-0.32)+1.30&<&{\rm log}([\rm O_{III}]/H\beta),\nonumber\\
0.73/({\rm log}([\rm O_{I}]/H\alpha)+0.59)+1.33&<&{\rm log}([\rm O_{III}]/H\beta)\nonumber,
\label{eq2}
\end{eqnarray}

while star-forming galaxies were selected using the selection criteria by Kauffmann
et al. (2003):

\begin{eqnarray}
0.61/({\rm log}([\rm N_{II}]/H\alpha)-0.05)+1.30&<&{\rm log}([\rm O_{III}]/H\beta),\nonumber\\
0.72/({\rm log}([\rm S_{II}]/H\alpha)-0.32)+1.30&<&{\rm log}([\rm O_{III}]/H\beta),\nonumber\\
0.73/({\rm log}([\rm O_{I}]/H\alpha)+0.59)+1.33&<&{\rm log}([\rm O_{III}]/H\beta), \nonumber
\end{eqnarray}

requiring a signal-to-noise ratio $\geq$ 6.
The galaxies residing in the area between the previous demarcation
lines (solid and dotted lines, respectively, in Figure~\ref{Fig1})
are identified as composite objects, which contain AGN as well as extended H$_{\rm II}$
regions. Our non-AGN sample consist of those galaxies excluding AGNs and composite
galaxies, without signal-to-noise cut.

To estimate the black hole mass ($M_{BH}$), we use the $M_{BH}$--stellar velocity dispersion
relation by Tremaine et al. (2002):

\begin{equation}
\label{BHmass}
{\rm log} (M_{\rm BH}/M_{\odot}) = 8.13 + 4.02 \times {\rm log} (\sigma/200~{\rm km~s}^{-1}),
\end{equation}

where a simple aperture correction to the stellar velocity dispersion was applied. Since velocity
dispersion measurements lower than the instrumental resolution are not robust, we will focus on AGNs
with $M_{\rm BH} > 10^{6.7}M_{\odot}$.

In order to discriminate between elliptical and disk galaxies, we used the prescription of
Park \& Choi (2005) in which early (ellipticals and lenticulars) and late (spirals) types are
segregated in a $u - r$ color versus $g - i$ color gradient space and in the concentration index space. 
They extensively tested the selection criteria through a direct comparison of visually assigned types
for a large sample of several thousand galaxies. The specific selection criteria can be found in Park \& Choi (2005). 
For more details the reader is referred to CPW09.

\section{Results}

Our Equation (\ref{LamObs}) requires the rotational velocity to calculate $\lambda$,
which is inferred from the absolute
magnitude in the $r$-band and using a TF relation (Pizagno et al. 2007).
To avoid the problem of internal 
absorption in edge-on galaxies (Cho \& Park 2009), and consequently underestimating rotational 
velocities, we limit the sample to spiral galaxies having seeing-corrected
isophotal axis ratios $b/a > 0.6$.

Once the galaxies in our sample are segregated into early- and late-type galaxies,
and into different activity classes such as normal, composite, and AGN hosting
galaxies, we computed $\lambda$ using Equation (\ref{LamObs}) for all
the late-type galaxies and obtained the spin distribution for the three
different subclasses: 41,662 normal, 2,273 composite and 1,026 AGNs.
Figure 2(a) shows the distribution
of $\lambda$ values for normal, composite, and AGN late-type host
galaxies. We can see a clear difference between the normal and the
AGN population, the former showing typically higher spin values with a
large dispersion compared with the latter, which is a more coherent
population of low spin galaxies.
The distribution of galaxies classified as composite follows almost exactly
the same distribution as AGN galaxies, probably due to the low number of
misclassified star-forming galaxies. Given that these galaxies are actually
a mixed group of AGN and non-AGN galaxies we will not consider them for the
rest of the analysis, and we will focus only in the difference between AGNs and
normal galaxies. 

Theoretical (Shaw et al. 2006), as well as empirical (Hernandez et al. 2007)
distributions of $\lambda$, are commonly described by a log-normal
function, in our case, the values describing the non-AGN and AGN distributions are, respectively,
$\lambda_{0}=0.059\pm0.006$, $\sigma_{\lambda}=0.400\pm0.005$ and $\lambda_{0}=0.042\pm0.004$, $\sigma_{\lambda}=0.350\pm0.003$,
showing a clear difference between the two populations.

We use the two sample Komogorov--Smirnov test to compare the distributions of normal and
AGN host galaxies to determine if they are drawn from the same parent distribution.
The result of the test
yields a negligible probability of $P \sim \times 10^{-180}$
that the two samples are drawn form the same parent distribution.

If we look at the $\lambda$ distribution of normal galaxies, splitting the sample
according to the absolute magnitude (Figure 2(b)),
we clearly see that the distributions get tighter and present
lower mean $\lambda$ values
for brighter galaxies, the same behavior we reported
in Cervantes-Sodi et al. (2008), where we found that massive galaxies tend to
have low spin dark matter halos and show low dispersion, while low
mass galaxies have higher $\lambda$ values and high dispersions around the median
value, a result later confirmed by Berta et al. (2008).
A striking difference appears for the case of AGN host galaxies
(Figure 2(c)), showing low mean $\lambda$ values
and tight distributions for all $M_{r}$. This result indicates
that only low spinning galaxies, at all mass ranges,
are suitable AGN hosts in the local universe.

Previous work has already pointed out that the recurrence of AGN activity is a
function of internal properties such as absolute magnitude, color, and velocity
dispersion. Kauffmann et al. (2003) state that massive galaxies with old stellar
populations frequently host AGNs and
CWP09 found that the fraction of AGN increases with galaxy luminosity,
and particularly for late-type galaxies,
this fraction increases for intermediate color systems,
with $u$ -- $r = 2\sim2.4$ and intermediate velocity dispersion ($\sim130$ km s$^{-1}$).
In this context, most of our AGN hosting galaxies are restricted to
having intermediate colors, a restriction that also constrains the
value of $\lambda$ given that color anticorrelates with $\lambda$ (Hernandez
\& Cervantes-Sodi 2006).
To make a fair comparison between AGN and non-AGN galaxies,
we constructed a control sample by selecting non-AGN galaxies
which match one to one the morphology, absolute magnitude, color and concentration
index of each galaxy in the AGN sample, guaranteeing that the
distributions of these parameters are the same for both populations.
Figure 2(d) shows how the typical $\lambda$ value of AGN hosting
galaxies tends to be lower than the one presented by the control sample
at all $M_{r}$ values. The uncertainties (throughout this paper) represent 1$\sigma$ confidence
intervals determined by the bootstrap resampling method.


\begin{figure}
\begin{tabular}{c}
\label{fractions}
\includegraphics[width=.475\textwidth]{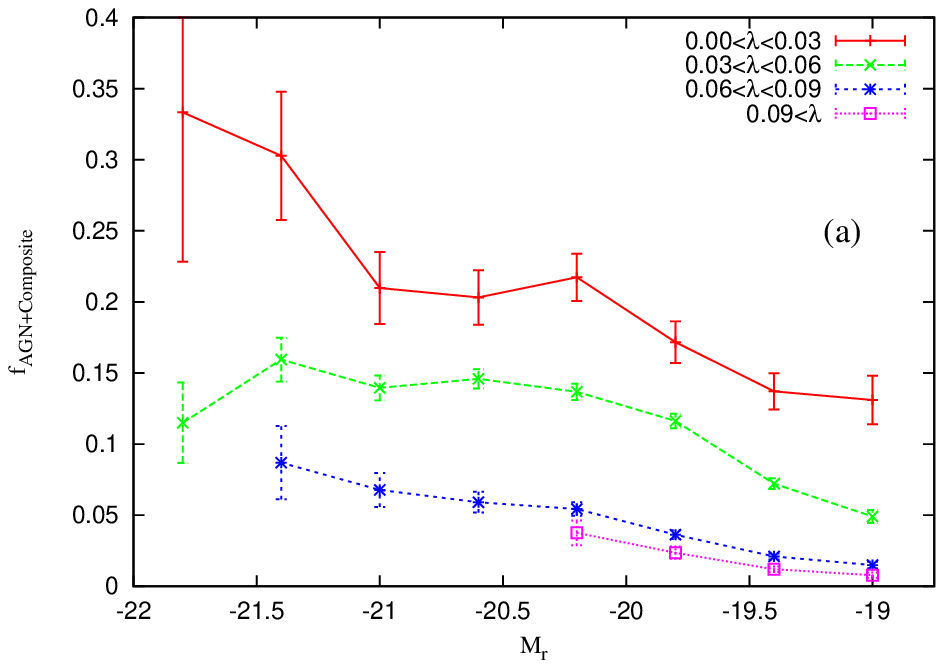} \\
\includegraphics[width=.475\textwidth]{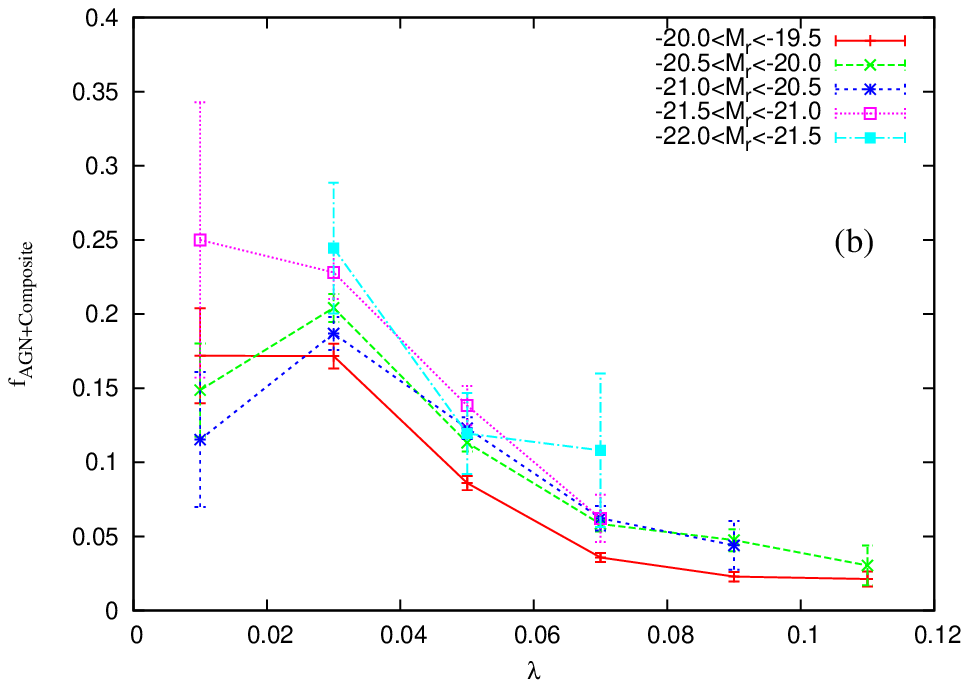}\\
\includegraphics[width=.475\textwidth]{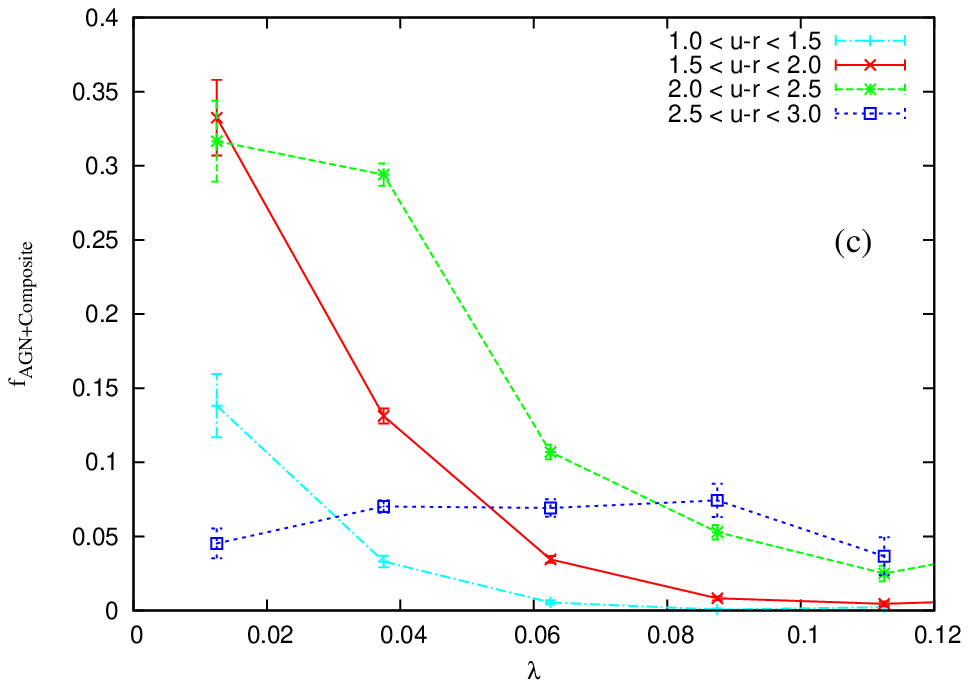}
\end{tabular}
\caption[ ]{AGN fraction as a function of(a) absolute magnitude with cuts of $\lambda$ 
and as a function of $\lambda$
with cuts in magnitude (b) and color (c).}
\end{figure}

A natural question arising from our previous results is
whether or not the AGN fraction depends on galactic spin.
Previous studies have shown the dependence of the AGN fraction
on the brightness of the parent galaxies (Mercurio et al. 2010, CWP09), or their associated
stellar or total mass (Pasquali et al. 2009). Figure 3(a)
shows the fraction of AGN+composite ($f_{\rm AGN+Composite}$)
galaxies as a function of absolute magnitude in different $\lambda$ cuts, 
where a moderate increase of $f_{\rm AGN+Composite}$ is present for increasing brightness,
while a large difference in this fraction is evident for the distinct $\lambda$ cuts.
Figure 3(b) gives a complementary study
for $f_{\rm AGN+Composite}$ as a function of galactic spin for thin $M_{r}$ slices,
where the dependence on $\lambda$ appears stronger than the case of $M_{r}$,
with all curves consistent with a single underlying dependence to within errors,
for all magnitude bins. This shows that the driving physical ingredient is
the value of $\lambda$ rather than magnitude.

CPW09 have already shown that late-type galaxies present weak dependence of $f_{\rm AGN+Composite}$
on magnitude or mass (accounted for through velocity dispersion) in addition to $u$ -- $r$ color, and
in the preceding section we found that the value of $\lambda$ is intimately linked to
the color of the galaxy in question. The $f_{\rm AGN+Composite}$ dependency on $\lambda$ for narrow
$u$ - $r$ bins is presented in Figure 3(c), where even at fixed color, a substantial
decrease of the AGN fraction results as $\lambda$ increases, at least in the
range of colors where the highest occurrence of AGNs is found. We see that regardless of mass or color, high
$\lambda$ systems will only very rarely present an AGN.


The tight correlations of $M_{\rm BH}$ with host galaxy properties, such as the
well-known one with the bulge luminosity (Kormendy \& Richstone 1995) or the tighter
$M_{\rm BH}$ versus bulge velocity dispersion relation (Graham et al. 2011),
have now been extended to more global
properties, not related to the bulge component alone.
Correlations with the spiral arm pitch angle of disk galaxies (Seigar et al.
2008), the mass of the dark matter halo (Ferrerase 2002; Booth \& Schaye
2010), and the total gravitational mass of the host galaxies (Bandara
et al. 2009) show how the correlations are not restricted only
to the bulge component and point to a SMBH--bulge--total-mass interrelation (Volonteri et al. 2011).
Given the important role the spin plays in establishing the morphology
and present-day structure of disk galaxies, and our previous result
that shows the direct influence of $\lambda$ on determining nuclear
activity, it is logical to expect a correlation between $\lambda$ and
$M_{\rm BH}$ for the active galaxies in our sample;
more specifically, an anticorrelation.

Let us consider the simple criterion by Larson (2010) for the formation of black holes,
where gas in a forming galaxy with a column density above a critical value $\Sigma_{\rm crit}$,
goes into forming a central black hole. If the gas in the forming galaxy is distributed
radially like the total mass with an isothermal density profile given by Equation
(\ref{RhoHalo}), with a gas to total mass ratio $f_{\rm g}$, the surface
density of gas at any radius $r$ in the disk will be $\Sigma_{\rm g}(r) =
f_{\rm g} V_{\rm d}^{2}/ 2 \pi G r$. For a rotationally supported disk, the scale
radius is given by $R_{\rm d} \simeq  \lambda r_{\rm vir}$,
where $r_{\rm vir}$ is the virial radius of the
system. This allow us to express the mass of gas within the critical column density criterion as

\begin{equation}
M_{g}(\Sigma < \Sigma_{\rm crit} )= \frac{f_{g} M^{2}}{\pi \Sigma_{\rm crit} (\lambda r_{\rm vir})^{2}}.
\label{MBH-lambda}
\end{equation}

This expression, that does not attempt to be more than a dimensional analysis estimate,
implies that at fixed total mass the mass of gas that could end up forming a black
hole is $M_{\rm BH} \propto \lambda^{-n}$, with $n=2$.
Theoretical studies report different values
for $n$ that fall in the range 1--4 (Colgate et al. 2003; Koushiappas et al.
2004), all of them establishing an anticorrelation between the mass of the
black hole and halo spin. Figure 4 (top panel) shows the empirical
dependence of the inferred $M_{\rm BH}$ with the spin parameter for our sample of AGN host
galaxies, showing the expected anticorrelation with $\lambda$, in this case with a best
fit described by log($M_{\rm BH}$) = ($-0.716 \pm 0.114$) log($\lambda$) $+ (6.455 \pm 0.155$).
In Figure 4 (bottom
panel) we present the result of splitting the sample according to the absolute magnitude of the
galaxies, where we can appreciate how at fixed $\lambda$ brighter (more massive) galaxies host
more massive black holes, with the anticorrelation with $\lambda$
present at all fixed $M_{r}$ cuts. In this case, the leading effect is the total
mass, with $\lambda$ introducing only a small but well-defined trend in the expected sense.

\begin{figure}
\centering
\begin{tabular}{c}
\label{l-mbh}
\includegraphics[width=0.475\textwidth]{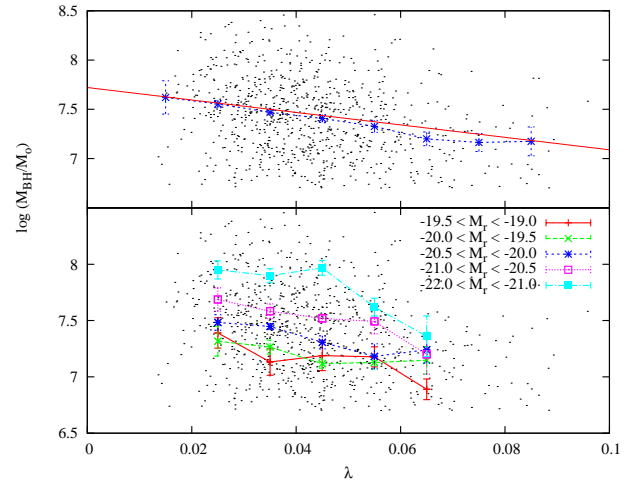}  
\end{tabular}
\caption[ ]{Upper panel:
$M_{\rm BH}$ versus $\lambda$ for AGNs hosted by late-type galaxies, showing
the median value for each bin. Lower panel: the same as in the upper panel
but showing the median $M_{\rm BH}$ values splitting the sample according to
the absolute magnitude.}

\end{figure}

\section{Conclusions}

Using an extensive sample of galaxies extracted from the SDSS DR5, we show that
the empirical distributions of the $\lambda$ spin parameter of AGNs and
non-AGN host galaxies are qualitatively and quantitatively different, the first distribution
having lower $\lambda$  values and smaller dispersion around the median
when compared with the latter. Another striking difference between the two populations is
that while the spin of normal galaxies presents a dependence on absolute magnitude, with
a systematic decrease of $\lambda$ with increasing luminosity, AGN host galaxies
present at all $M_{r}$ low $\lambda$ values. This result, in addition to the increase
of the AGN fraction for decreasing $\lambda$, highlights the requirement
for galaxies to have low spin in order to host AGNs and should be taken into account
in semi-analytic models that exclusively use the halo or stellar mass to reproduce
the fraction of galaxies belonging to different activity classes (e.g., Wang \& Kauffmann 2008;
Fontanot et al. 2011). Seeding AGNs into modeled dark matter halos should primarily consider
$\lambda$, in order to better model the observed universe.

For the AGN sample we found that the inferred mass of the SMBHs shows a weak dependence on
the spin of the hosting galaxies, with increasing mass for decreasing spin, a logical
result if we need the gas to be accreted onto the central massive object,
a process that can be more effectively accomplished if the raw material originally
has low angular momentum. We recovered the leading $M_{\rm BH}$-$\lambda$ relation at fixed
absolute magnitude, with typically higher SMBH mass for brighter galaxies at fixed
$\lambda$, showing a double dependence of $M_{\rm BH}$ on $M_{r}$ and $\lambda$.
Given that the trend with the spin is present at all $M_{r}$ bins, this could
account, to a certain degree, for the dispersion in the well established
$M_{\rm BH}$--luminosity relation.

We thank the anonymous referee for a thorough reading and constructive comments that helped to
enrich our study. B.C-S. thanks Cheng Li for useful comments.
Y.Y.C. was supported by grant from the Kyung Hee University (KHU-20100179)
and the National Research Foundation of Korea to the Center for Galaxy Evolution Research.

\end{document}